# Microscopic Structural Study on the Growth History of Granular Heaps Prepared by the Raining Method

Hanyu Li,[1] Houfei Yuan,[3] Zhikun Zeng,[3] Shuyang Zhang,[3] Chijin Zhou,[1] Xinyu Ai,[1] and Yujie Wang[1,2,3,*]

[1]Department of Physics, College of Mathematics and Physics, Chengdu University of Technology, Chengdu 610059, China
[2]State Key Laboratory of Geohazard Prevention and Geoenvironment Protection, Chengdu University of Technology, Chengdu 610059, China
[3]School of Physics and Astronomy, Shanghai Jiao Tong University, Shanghai 200240, China

Granular heaps are critical in both industrial applications and natural processes, exhibiting complex behaviors that have sparked significant research interest. The stress dip phenomenon observed beneath granular heaps continues to be a topic of significant debate. Current models based on force transmission often assume that the packing is near the isostatic point, overlooking the critical influence of internal structure and formation history on the mechanical properties of granular heaps. Consequently, these models fail to fully account for diverse observations. In this study, we experimentally explore the structural evolution of three-dimensional (3D) granular heaps composed of monodisperse spherical particles prepared using the raining method. Our results reveal the presence of two distinct regions within the heaps, characterized by significant differences in structural properties such as packing fraction, contact number, and contact anisotropy. We attribute these structural variations to the differing formation mechanisms during heap growth. Our findings emphasize the substantial influence of the preparation protocols on the internal structure of granular heaps and provide valuable insights into stress distribution within granular materials. This research may contribute to the development of more accurate constitutive relations for granular materials by informing and refining future modeling approaches.

## I. INTRODUCTION

Granular heaps play a vital role in numerous industrial applications and natural phenomena [1-3]. Despite their seemingly simple structure, granular heaps exhibit a wealth of intricate behaviors and complex features that have attracted recurring interest over the years. These include avalanching behavior [4,5], which is closely related to the well-known concept of self-organized criticality (SOC) [6], the distinctive surface topography of sand dunes in deserts [7,8], various factors influencing the angle of repose [9-11], and the stress distribution underneath the pile [12-14]. Specifically, the stress dip observed underneath granular heaps has sparked significant debate regarding its origin. Edwards and coworkers proposed that there exists asymmetry in the supporting forces exerted on a particle from its supporting particles, leading forces to propagate along chain-like arch structures, which effectively directs stress sideways to reduce the load underneath [15]. Building on this simple idea, the scalar $q$ model [16] and the more complex oriented stress linearity (OSL) model [12] have been developed. In contrast to conventional elasto-plastic soil mechanics models [17,18], these models attempt to close the constitutive relations by stress components only without strain. This approach is motivated by the fact that granular particles are rigid and close to the hard-sphere limit [19], rendering elastic energy negligible when deformations are small. By further assuming that granular systems are nearly isostatic, this new closure scheme leads to a hyperbolic equation for force transmission instead of an elliptic one [20]. This prediction is used to explain the appearance of arches or force chain structures in granular materials and the potential hyperbolic instead of ordinary elliptical force responses, although the experimental verification remains controversial [21]. Concurrently, numerical simulations by Zheng and Yu [22] demonstrated that the appearance of the stress dip depends on the heap preparation history, indicating that its appearance is non-universal and protocol-dependent. This underscores the need for studies that account for both the specific preparation protocol and history of granular piles to fully understand their properties. Topić *et al.* performed simulations to investigate the density distribution and angle of repose in three-dimensional (3D) heaps of monodisperse spherical particles [23]. Through microscopic structural analysis, their analysis revealed that when the particles are deposited from a small outlet, the resulting heaps exhibit three distinct angles of repose (including an internal angle) and four characteristic density zones, all closely tied to the deposition history. Conversely, a large outlet size results in the merging of the two distinct surface repose angles. These findings further underscore the critical role of deposition protocol and history in determining the properties of granular heaps. Nevertheless, prior experimental studies on granular heaps have primarily focused on macroscopic properties, such as external angles of repose or base stress distributions, often lacking detailed microscopic information into the internal structure and force distribution. Although photo-elastic experiments offer valuable structural and force information, they are restricted to two dimensions (2D) [21]. Therefore, there is a pressing need for systematic experimental investigations of the microscopic structure of 3D granular piles that consider their growth history.

In this paper, we present X-ray tomography experiments to investigate the structural evolution and growth history of 3D heaps composed of monodisperse spherical particles. We investigate the packing fraction, number of contacts, and contact anisotropy as a function of heap growth. Our results reveal that the internal structure of a granular heap can be divided into two distinct regions, each characterized by specific structural features and contact properties that emerge during heap formation and growth.

## II. EXPERIMENTAL METHODS

We prepare granular heaps by depositing approximately 30,000 monodisperse spherical acrylonitrile butadiene styrene plastic (ABS) beads, each with a diameter of $d = 3$ mm, into a plexiglass cylindrical container with an inner diameter of $60d$. Following the standard raining method [24], the beads are released from a source with an outlet diameter $7d$, positioned directly above the container's center at a height of approximately $32d$ from the bottom. A sieve with a mesh size of 5 mm is placed just below the outlet to ensure a homogeneous "rain" of particles. The initial state of the heap is formed by depositing approximately 8,600 particles, forming a conical heap with a height of approximately $7d$. Subsequently, five additional batches, each containing approximately 4,300 particles, are added bringing the final heap height to approximately $17d$. This methodology allows for a systematic investigation of six distinct growth stages of the heap. After each deposition, 3D structures of the packing are obtained through X-ray tomography scans using a medical CT scanner (UEG Medical Group Ltd., 0.2 mm spatial resolution). Using image processing procedures similar to those described in our previous studies [25,26], we obtain the centroid coordinates of each particle with an error of less than $3 \times 10^{-3} d$. Results for each configuration are averaged over 5 independent realizations to enhance statistical reliability. To mitigate boundary effects, only particles at least

$2d$ away from the heap boundary are included in the analysis. For comparison, granular packings with a wide range of packing fractions are generated through mechanical tapping at different tapping intensities. Further details on the tapping experiments can be found in Ref. [27].

## III. RESULTS

### A. Two regions within the heap distinguished by packing density

To investigate the structural evolution of the granular heaps during growth, we first examine the azimuthally averaged local volume fraction $\phi = \langle v_p \rangle / \langle v \rangle$, where $v_p$ and $v$ represent the volumes of the particles and their associated Voronoi cells, respectively. Using cylindrical coordinates $(R, H, \psi)$ coaxial with the heap, as shown in Fig. 1(b), the azimuthal average is calculated over different azimuthal angle $\psi$ for all particles with identical $R$ and $H$. This averaging is employed due to the conical shape and circular symmetry of the three-dimensional granular heaps in each horizontal plane, with the apex as the center.

Figure 1(a) shows the spatial distributions of $\phi$ as a function of radial distance $R$ and height $H$ for six packings, documenting the heap's growth history. The data are coarse-grained by averaging $\phi$ of particles within square blocks of size $d^2$. Among these packings, particles located directly below the outlet, within a cylindrical region of diameter approximately $7d$, typically have higher $\phi$ compared to those farther away from the center of the heap. Then we delineate the heap into two regions using the contour line of $\phi = 0.62$: region $A$ at the center and region $B$ surrounding it [Fig. 1(b)]. Similar to the findings of Topić et al. [23], we observe that the heaps exhibit an external angle of repose $\gamma \approx 25°$ along with a larger internal angle of repose $\alpha \approx 54°$, defined by the boundary between regions *A* and *B* in Fig. 1(a). Unlike the external angle of repose, which primarily depends on particle friction, the internal angle of repose is often recognized as being closely related to the preparation history of the heap [23]. Figure 1(c) shows the evolution of $\phi$ for particles in both regions as a function of deposition number *n*. It is important to note that *n* = 1, 2, …, 6 represents the entire heap after the *n*-th deposition, whereas *m* = 1, 2, …, 6 refers only to the layers of particles added during the *n*-th deposition. We observe a slight decrease in volume fraction $\phi$ at the bottom of the heap, i.e., in $A_{m=1}$, across different growth stages, while $\phi$ remains relatively constant in other parts of the heap. Additionally, a significant decrease in $\phi$ with increasing radial distance *R* is noted. This behavior is attributed to the distinct formation mechanisms in the two regions, which correspond to varying intensities of energy dissipation. For region *A*, beads fall directly from the outlet with substantial kinetic energy and collide with particles below to form new layers. Particles deposited withstand the impact of newly arrived beads and facilitate their settling, forming more stable structures, *i.e.*, higher $\phi$. This finding aligns with the empirical observations from the pluviation protocol, a standard particle packing preparation protocol in soil mechanics, which suggests that particles falling from greater heights with higher kinetic energy tend to achieve denser packing [28-30]. In contrast, particles in region *B* originate from those already flowing down the heap, having undergone at least one inelastic collision with other particles. As these beads descend further down the slope, they gain kinetic energy from gravity, yet this energy is still reduced due to inelastic collisions and frictional dissipation. Consequently, beads deposited farther from the center tend to form looser structures, consistent with experimental observations. Our results thus emphasize the critical role of kinetic energy in the growth of granular heaps, an aspect often overlooked in previous studies [31].

### B. Contact number and contact anisotropy

To further elucidate the structural characteristics of the granular heaps, we calculate the average contact number $Z$ and the fabric intensity $\beta$ of particles in different regions. Ideally, particles in direct contact should have zero surface-to-surface distance. However, due to experimental uncertainties, such as finite X-ray spatial resolution and inaccuracies in determining particle sizes and centroids, particles may be erroneously identified as either overlapping or having gaps. To address this issue, we employ standard procedures for contact determination, using a complementary error function fitting approach [32]. The contact network is further characterized by fabric tensor [33,34]:

$$\mathbf{A} = \frac{1}{N_c}\sum_{a=1}^{N_c}\mathbf{n}^a \otimes \mathbf{n}^a - \frac{1}{3}\mathbf{I}, \tag{1}$$

where $N_c$ is the total number of contacts, $\mathbf{n}$ is the unit contact vector between the centers of two contacting particles, and $\mathbf{I}$ is the unit tensor. The contact anisotropy is quantified by the ratio of the maximum and minimum eigenvalue of $\mathbf{A}$, referred to as the fabric intensity $\beta$. The distributions of $Z$ and $\beta$ for the growing heap are shown in Figs. 2(a) and 3(a), while their evolution in regions $A$ and $B$ across different deposition stages is depicted in Figs. 2(b) and 3(b). The contact number $Z$ exhibits a similar evolution trend to the volume fraction $\phi$, with significantly higher values in region $A$ compared to region $B$. This observation reinforces the hypothesis that the packing structures in region $A$ are more stable than those in region $B$. Notably, regions of the heap with higher $\phi$ also exhibit larger fabric intensity $\beta$, which contrasts with the behavior observed in packings prepared by mechanical tapping [see Fig. 4(a)].

Nevertheless, the Voronoi cell volume distributions $P(v_p)$ of the packings, prepared using different methods but exhibiting similar $\phi$, are nearly identical [inset of Fig. 4(a)]. This indicates that, despite variations in local contact structures, these systems are thermodynamically equivalent within the Edwards volume ensemble [34]. This further implies that the physics at the contact and particle scales are weakly coupled in these systems, distinguishing them from those near isostaticity.

### C. History-dependent spatial distribution of contact points

To investigate the disparity in contact distributions, we compute the spatial distribution of contact points on the particle surface, denoted as $f_i(\mathbf{r}) = \delta(\mathbf{r} - \mathbf{r}_j)$, where $\mathbf{r}_j(r,\theta,\varphi)$ represents the contact point position in spherical coordinates defined for bead $i$. For averaging purposes, we define $\varphi = 0$ as the direction from the particle center to the heap's center in the spherical coordinates. The function $f_i(\mathbf{r})$ is normalized by $\iint_{S_i} f_i(\mathbf{r})ds = Z_i$, where $S_i$ is the surface area of bead $i$ in units of $d^2$, and $Z_i$ is its contact number. $\langle f(\mathbf{r}) \rangle$ represents the average of $f_i(z)$ across all particles in region $A$ or $B$ [Figs. 4(c)-4(h)]. The front views of $\langle f(\mathbf{r}) \rangle$ in Figs. 4(c)-4(h) are taken at $\varphi = 90°$, with the left-hand side oriented towards the heap center. The probability distribution functions (PDFs) of contact points $p(\theta)$ for particles in regions $A$ and $B$ are shown in Fig. 4(b). For comparison, we also calculate $p(\theta)$ for packings obtained from vertical tapping experiments with $\phi_1 = 0.605$ and $\phi_4 = 0.631$ [blue and red curves in Fig. 4(b)] and analyze $\langle f(\mathbf{r}) \rangle$ for packings with $\phi_1 = 0.605$, $\phi_2 = 0.612$, $\phi_3 = 0.626$, and $\phi_4 = 0.631$ [Figs. 4(e)-4(h)]. In regions $A$ and $B$, we observe that contact points tend to concentrate at the top and bottom of the particles, with fewer contacts distributed near the equator. A similar pattern emerges in tapping experiments with high tap intensity $\Gamma$, i.e., low $\phi$. This phenomenon is likely due to vertical contacts offering greater mechanical stability under vertical impact compared to horizontal contacts. However, in tapped packings, as $\Gamma$ decreases, the contacts become more uniformly distributed across the particle surface [Figs. 4(e)-4(h)]. This occurs because, at lower $\Gamma$, particles have lower kinetic energy, which promotes the formation of stable configurations governed by lateral frictional forces, thereby leading to a more uniform distribution of contacts. Note that in region *A* of the heap, despite with $\phi$ similar to that of Fig. 4(g), the contact distributions are significantly different. This is most likely due to the preparation method. Unlike the tapped packings, particles in the raining method are deposited sequentially. As a result, collective structures like bridges, which rely on mutual support, are less likely to form during the process. Instead, to withstand the vertical impacts, contacts are more likely to distribute near the poles. Overall, this suggests that the contact structure is highly dependent on the preparation protocol and does not exhibit a one-to-one correspondence with $\phi$. Additionally, in region *B*, the "blue band", which denotes fewer contacts near the particle equator, appears slightly tilted [Fig. 4(d)]. This occurs because, during heap formation, the existing structures are formed to withstand particles flowing down the slope, leading them to form contacts and stabilized structures that are tilted to accommodate this specific impact geometry. This further emphasizes the strong connection between the internal contact structures of the heap and its growth history.

## IV. CONCLUSION

In summary, we use X-ray tomography to investigate the internal structure and growth history of 3D granular heaps composed of monodisperse spherical particles prepared via the raining method. Our analysis of the spatial distribution of the packing fraction reveals that these heaps can be characterized by two distinct regions, *A* and *B*, with clear differences in structural properties such as the contact number and contact anisotropy. These regions emerge due to different formation mechanisms during heap growth, as local structures are formed to resist the impact of impinging particles. The contrasting internal structures of heaps prepared by the raining and tapping methods underscore the strong dependence of granular packing configurations on the preparation protocol. Returning to the controversy surrounding the stress dip underneath the heap, our study clearly demonstrates that preparation history is pivotal in determining the final force structure of the system. Consequently, the oversimplified assumption that such systems are isostatic fails to account for such diverse observations. Moreover, it is experimentally observed that granular packings typically do not reside at the isostatic point [27]. Nevertheless, in our opinion, the debate between force transmission models and the traditional elasto-plastic soil mechanics models remains far from resolved. Notably, even when systems are far from the isostatic, collective structures such as bridges can still form. This supports Edwards' original proposal that constitutive relations could be closed using stress components alone, as his explanation for the occurrence of force chains does not require isostaticity. Additionally, it remains valid that elastic energy plays a negligible role in granular materials at low pressure, highlighting the need for developing constitutive models that are independent of strain. Indeed, as suggested by a recent study [35], the equilibrium states of the granular packings can, in principle, be determined using an Edwards-type thermodynamic free energy, where the system's rigidity emerges from entropic forces rather than elastic energy. In this framework, a constitutive relation without strain is possible, and this approach does not necessitate isostaticity. This may ultimately provide a resolution to this issue.

ACKNOWLEDGMENTS

The work is supported by the National Natural Science Foundation of China (No. 12274292).

Corresponding author

*yujiewang@sjtu.edu.cn

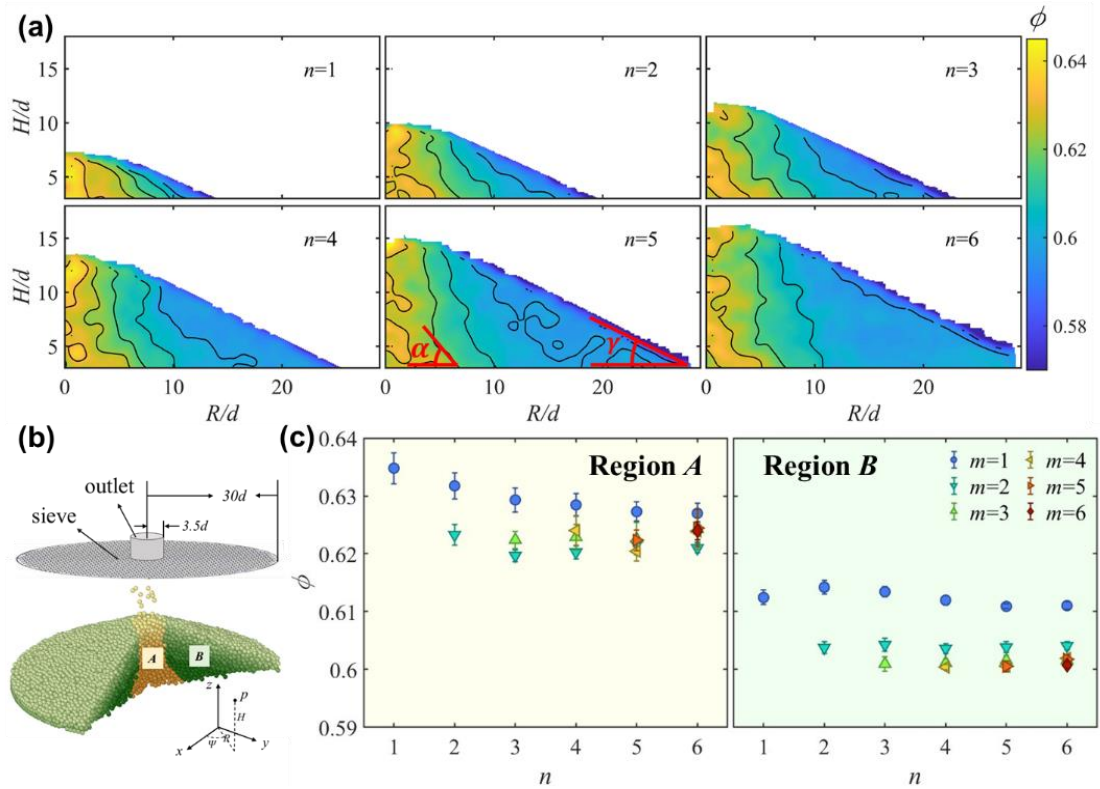

FIG. 1. (a) Spatial distributions of $\phi$ as a function of radial distance $R$ and height $H$ for different deposition numbers $n$. (b) Schematic diagram of the granular heap experiment. The yellow area represents region $A$, and the green area represents region $B$. Different shadows represent the different depths of the particles within the heap, with particles closer to the bottom of the pile corresponding to deeper shadows. (c) Volume fraction $\phi$ as a function of deposition number $n$ in region $A$ and region $B$, respectively. Different markers represent different layers $m$.

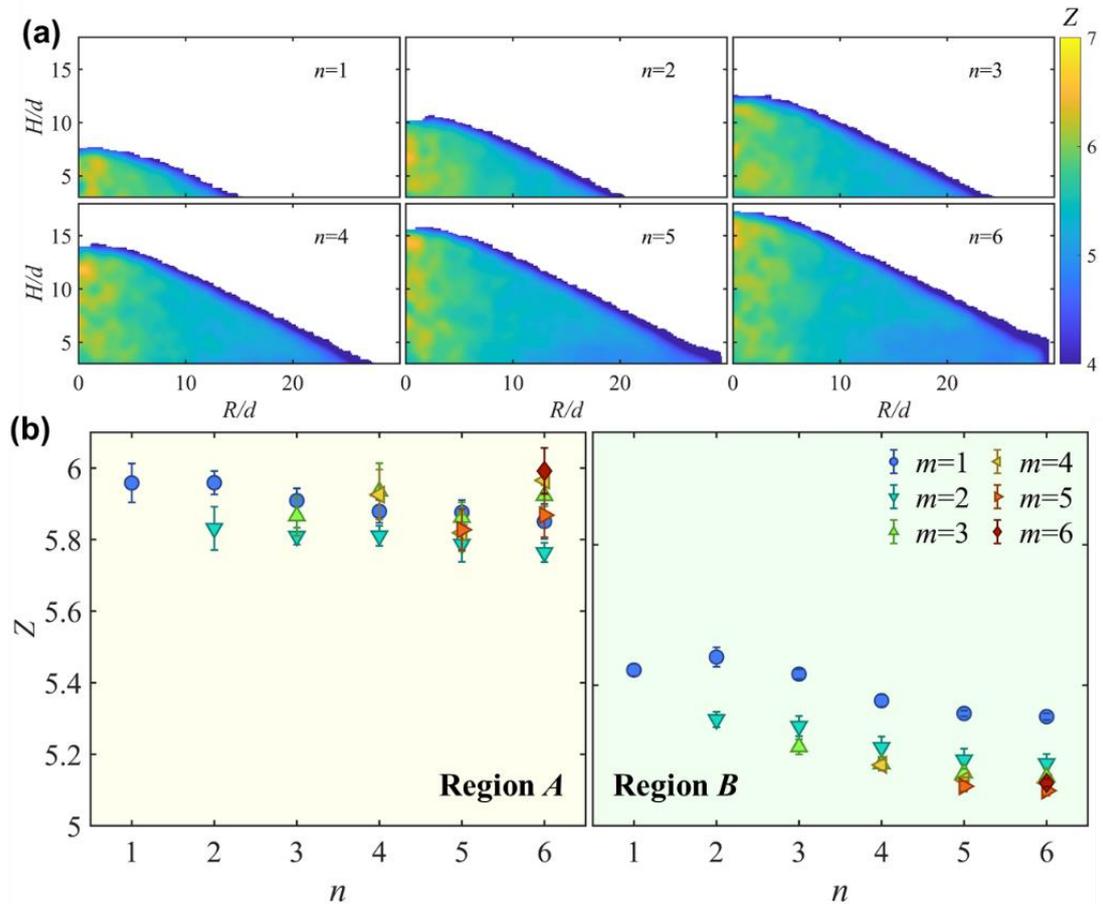

FIG. 2. (a) Spatial distributions of the contact number $Z$ as a function of radial distance $R$ and height $H$ for different deposition numbers $n$. (b) The contact number $Z$ as a function of deposition number $n$ for different layers $m$ in region $A$ and region $B$, respectively.

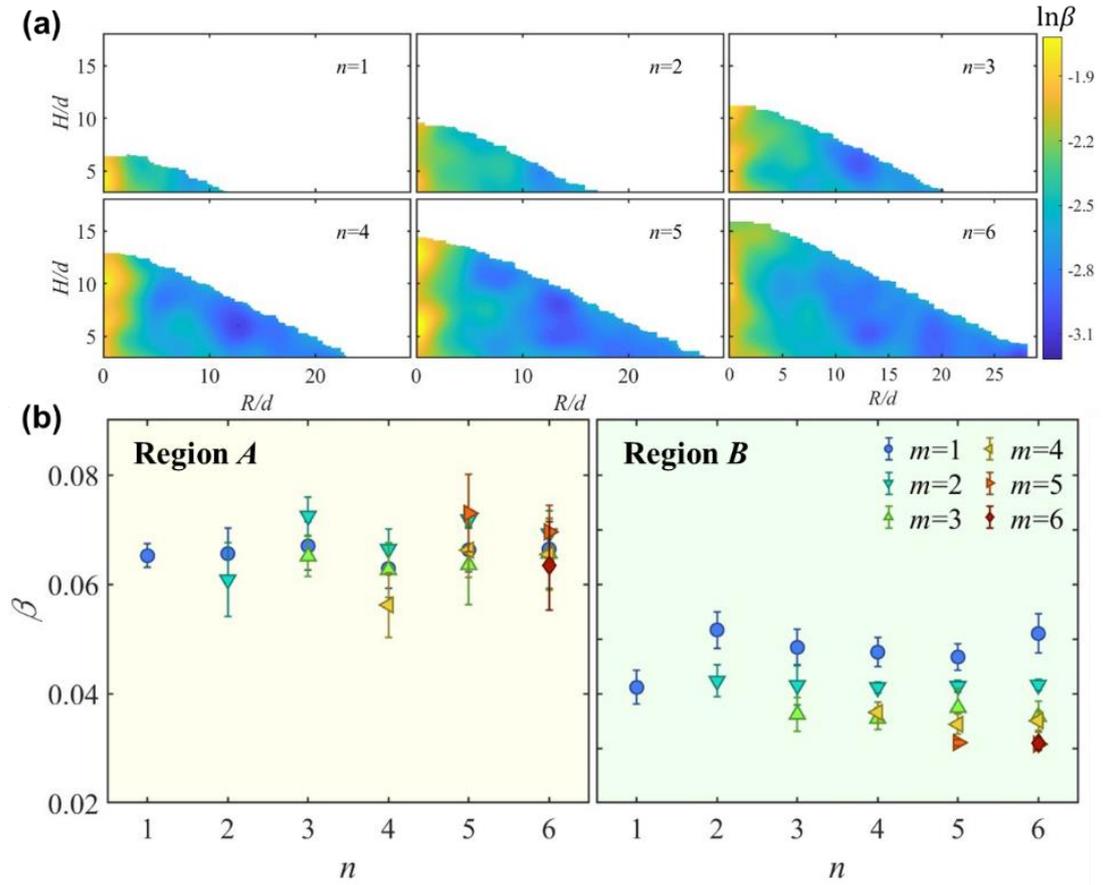

FIG. 3. (a) The spatial distributions of the fabric intensity $\beta$ as a function of radial distance $R$ and height $H$ for different deposition number $n$. (b) Fabric intensity $\beta$ as a function of deposition number $n$ for different layers $m$ in region $A$ and region $B$, respectively.

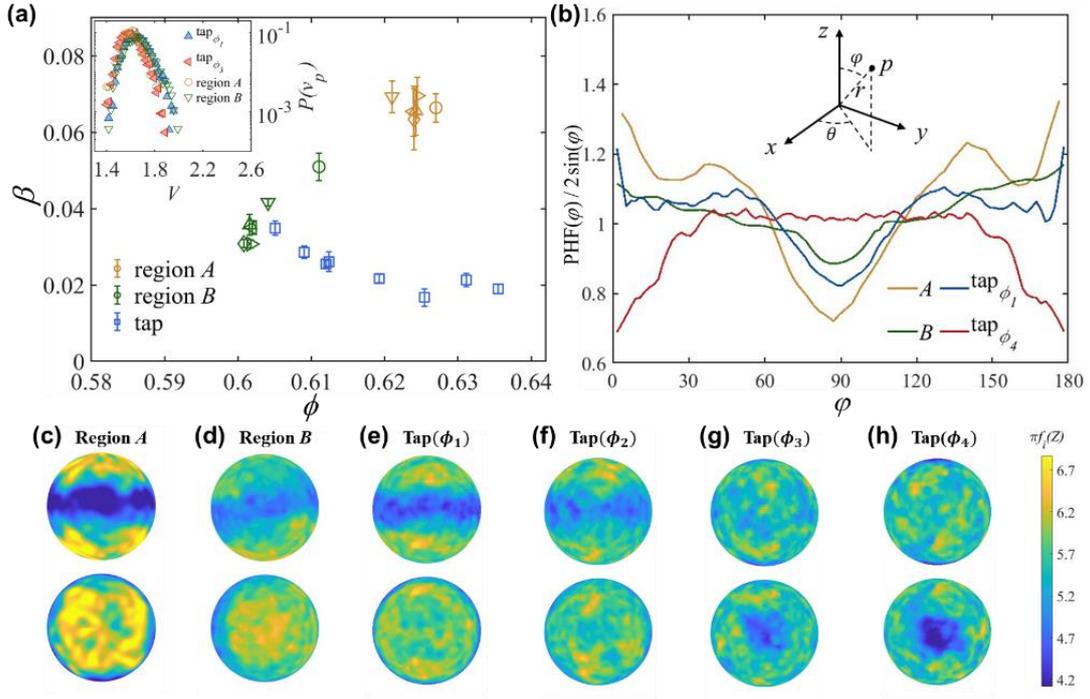

FIG. 4. (a) Fabric intensity $\beta$ as a function of volume fraction $\phi$ in region $A$, region $B$ and tapped packings. Inset: $P(v_p)$ of Voronoi cell volume in region $A$, $B$, and tapped packings with $\phi_1 = 0.605$ and $\phi_3 = 0.626$. (b) PDFs of contact points $PDF(\varphi)$ as a function of $\varphi$ for particles in regions $A$, $B$ and tapped packings with $\phi_1 = 0.605$ and $\phi_4 = 0.636$. (c-h) The spatial distributions of $f_i(z)$ on the particle surface in region $A$, $B$, and tapped packings with $\phi_1 = 0.605$, $\phi_2 = 0.612$, $\phi_3 = 0.626$ and $\phi_4 = 0.631$. The first line (second line) displays the sectional view (bottom view). The spatial distribution $f_i(z)$ of contact points is normalized by the spherical surface area.